\documentclass[epj,nopacs]{svjour}
\usepackage{graphics,times}
\newcommand{\Li}[1]{\mathop{\mathrm{Li}}\nolimits_{#1}}

\begin{document}
\title{HQET heavy--heavy vertex diagram with two velocities}
\author{A.G.~Grozin\inst{1}\thanks{Permanent address:
Budker Institute of Nuclear Physics, Novosibirsk, Russia}\and
A.V.~Kotikov\inst{2}}
\institute{Institut f\"ur Theoretische Teilchenphysik,
Karlsruher Institut f\"ur Technologie, Germany;
\email{A.G.Grozin@inp.nsk.su}\and
Bogoliubov Laboratory of Theoretical Physics, JINR, Dubna, Russia;
\email{kotikov@theor.jinr.ru}}
\date{}
\abstract{
The one-loop HQET heavy--heavy vertex diagram with arbitrary
powers of all three denominators and arbitrary residual energies
is investigated.
Various particular cases in which the result becomes simpler
are considered.}
\maketitle

\section{Introduction}
\label{Intro}

The heavy--heavy quark current in HQET (see, e.g., \cite{N:94,G:04})
transforms a heavy quark with velocity $v_1$
into a heavy quark with velocity $v_2$.
Loop diagrams with a velocity-changing vertex appear, e.g.,
when the anomalous dimension of the heavy-heavy current~\cite{KR:87}
or correlators involving this current (see~\cite{N:94})
are considered.
They can be reduced to a set of master integrals.
Some of them have the form of the one-loop vertex diagram
with various $\varepsilon$-dependent powers of the denominators.

\begin{figure}
\begin{center}
\begin{picture}(87,25)
\put(21,11){\makebox(0,0){\includegraphics{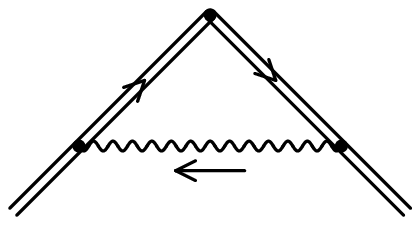}}}
\put(21,4.5){\makebox(0,0)[t]{$k$}}
\put(13,14){\makebox(0,0)[br]{$k+p_1$}}
\put(29,14){\makebox(0,0)[bl]{$k+p_2$}}
\put(66,11){\makebox(0,0){\includegraphics{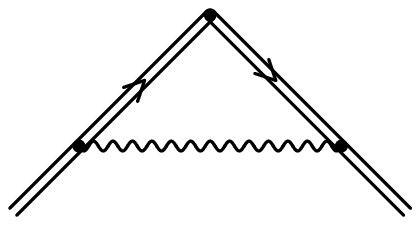}}}
\put(66,23){\makebox(0,0)[b]{0}}
\put(51,7){\makebox(0,0)[br]{$-v_1 t_1$}}
\put(81,7){\makebox(0,0)[bl]{$v_2 t_2$}}
\end{picture}
\end{center}
\caption{The one-loop HQET heavy--heavy vertex diagram
in momentum and coordinate space}
\label{Diagram}
\end{figure}

Here we consider this one-loop diagram (Fig.~\ref{Diagram})
in dimensional regularization in the most general case.
It is
\begin{eqnarray}
&&I(n_1,n_2,n_3;\vartheta;\omega_1,\omega_2) = \frac{1}{i\pi^{d/2}}
\int \frac{d^d k}{D_1^{n_1} D_2^{n_2} D_3^{n_3}}\,,
\nonumber\\
&&D_1 = - 2 (k + p_1) \cdot v_1\,,\quad
D_2 = - 2 (k + p_2) \cdot v_2\,,
\nonumber\\
&&D_3 = - k^2\,.
\label{Definition}
\end{eqnarray}
It depends on the residual energies $\omega_{1,2}=p_{1,2}\cdot v_{1,2}$
and the Minkowski angle $\cosh\vartheta=v_1\cdot v_2$.
It is symmetric:
\begin{equation}
I(n_1,n_2,n_3;\vartheta;\omega_1,\omega_2) =
I(n_2,n_1,n_3;\vartheta;\omega_2,\omega_1)\,;
\label{Symmetry}
\end{equation}
in the following, we shall not explicitly write down relations
obtained by this obvious symmetry.
This diagram vanishes at integer $n_3\le0$.
For integer $n_2\le0$,
it becomes the self-energy diagram with a numerator;
if, in addition, $n_1$ is integer and $n_1\le0$, the diagram vanishes.
The diagram has cuts in $\omega_{1,2}$ from 0 to $+\infty$.
We shall consider $\omega_{1,2}<0$,
other cases can be treated by analytical continuation.

In physical applications, $n_{1,2}=m_{1,2}+2l_{1,2}\varepsilon$,
$n_3=m_3+l_3\varepsilon$, where $m_i$, $l_i$ are integer, and $l_i\ge0$.
All integrals with given $l_i$ can be reduced,
using integration by parts~\cite{CT:81}, to 3 master integrals
\begin{eqnarray}
&&I(1+2l_1\varepsilon,1+2l_2\varepsilon,1+l_3\varepsilon;\vartheta;\omega_1,\omega_2)\,,
\nonumber\\
&&I(2l_1\varepsilon,1+2l_2\varepsilon,1+l_3\varepsilon;\vartheta;\omega_1,\omega_2)\,,
\nonumber\\
&&I(1+2l_1\varepsilon,2l_2\varepsilon,1+l_3\varepsilon;\vartheta;\omega_1,\omega_2)\,.
\label{Masters}
\end{eqnarray}
If $l_1=0$, the second integral is trivial:
\begin{eqnarray}
&&I(0,n_2,n_3;\vartheta;\omega_1,\omega_2) = I(n_2,n_3) (-2\omega_2)^{d-n_2-2n_3}\,,
\nonumber\\
&&I(n_1,n_2) = \frac{\Gamma(n_1+2n_2-d) \Gamma(d/2-n_2)}{\Gamma(n_1) \Gamma(n_2)}\,,
\label{I1}
\end{eqnarray}
see~\cite{BG:91} (similarly for the third integral if $l_2=0$).
Reduction algorithms for all cases has been constructed
and implemented in \texttt{REDUCE},
and can be downloaded from~\cite{Red}.

Using HQET Feynman parametrization (see, e.g., \cite{G:04}),
we can write the diagram~(\ref{Definition})
as an integral in 2 parameters $y_{1,2}$
which have dimensionality of energy and vary from 0 to $\infty$.
After the substitution $y_1=yx$, $y_2=y(1-x)$,
the integral in $y$ can be calculated:
\begin{eqnarray}
&&I(n_1,n_2,n_3;\vartheta;\omega_1,\omega_2) = I(n_1+n_2,n_3)
\frac{\Gamma(n_1+n_2)}{\Gamma(n_1) \Gamma(n_2)}
\nonumber\\
&&{}\times
\int_0^1 A^{-l} (-2\Omega)^{-r} x^{n_1-1} (1-x)^{n_2-1} dx\,,
\nonumber\\
&&A = x^2 + (1-x)^2 + 2 x (1-x) \cosh\vartheta
\nonumber\\
&&\quad{} = \left[1 - (1-e^\vartheta) x\right]
\left[1 - (1-e^{-\vartheta}) x\right]\,,
\nonumber\\
&&\Omega = \omega_1 x + \omega_2 (1-x)\,,
\nonumber\\
&&l = \frac{d}{2} - n_3\,,\quad
r = n_1 + n_2 - 2 l\,.
\label{Main}
\end{eqnarray}
This integral depends on $d$ and $n_3$ only via $l$;
therefore, shifting $d$ by $\pm2$~\cite{T:96}
is equivalent to shifting $n_3$ by $\pm1$.
It is Lauricella function $F_D$~\cite{S:66}
\begin{eqnarray}
&&I(n_1,n_2,n_3;\vartheta;\omega_1,\omega_2)
= I(n_1+n_2,n_3) (-2\omega_2)^{-r}
\nonumber\\
&&{}\times F_D(n_1;l,l,r;n_1+n_2;1-e^\vartheta,1-e^{-\vartheta},\xi_-)\,,
\label{Lauricella}
\end{eqnarray}
where
\begin{equation}
\xi_\pm = 1 \pm \xi\,,\quad
\xi = \frac{\omega_1}{\omega_2}\,.
\label{xi}
\end{equation}
This result can also be obtained in coordinate space
(Fig.~\ref{Diagram}).
Substituting $t_1=tx$, $t_2=t(1-x)$,
we can calculate the integral in $t$ and reproduce~(\ref{Main}).
Another one-dimensional integral representation can be obtained
by separating the $k$ space into the 2-dimensional longitudinal
and $(d-2)$-dimensional transverse subspaces~\cite{K:93},
but this representation seems more complicated.%
\footnote{In the ordinary relativistic theory,
the one-loop vertex diagram in the most general case is given by $F_D$~\cite{K:11};
the massless one via Appell $F_4$ functions~\cite{BD:87};
and the one with any masses but all powers of the denominators equal to 1
via Appell $F_1$~\cite{FJT:03}.}

The case $v_1=v_2$ ($\vartheta=0$) has been considered in~\cite{BBG:93}:
\begin{eqnarray}
&&I(n_1,n_2,n_3;0;\omega_1,\omega_2)
= I(n_1+n_2,n_3) (-2\omega_2)^{-r}
\nonumber\\
&&{}\times{}_2F_1\left(\left.
\begin{array}{c}
n_1,r\\
n_1+n_2
\end{array}
\right| \xi_- \right)\,.
\label{BBG}
\end{eqnarray}
If $n_{1,2}$ are integer, it reduces to trivial cases~(\ref{I1}).

In the single-scale case $\omega_1=\omega_2$,
the integral~(\ref{Main}) gives the Appell function $F_1$~\cite{S:66}
\begin{eqnarray}
&&I(n_1,n_2,n_3;\vartheta;\omega,\omega)
= I(n_1+n_2,n_3) (-2\omega)^{-r}
\nonumber\\
&&{}\times
F_1(n_1;l,l;n_1+n_2;
1-e^\vartheta,1-e^{-\vartheta})\,.
\label{Appell}
\end{eqnarray}
Also,
\begin{equation}
I(n_1,n_2,n_3;\vartheta;\omega,0)
= I(d-n_2-2n_3,n_2,n_3;\vartheta;\omega,\omega)
\label{omega0}
\end{equation}
(similarly for $\omega_1=0$);
so, these cases reduce to $\omega_1=\omega_2$.

\section{Exact results}
\label{Exact}

In the Euclidean region $\vartheta=i\vartheta_E$, let's consider
\begin{equation}
I = \frac{I(n_1,n_2,n_3;i\vartheta_E;\omega_1,\omega_2)}%
{I(n_1+n_2,n_3) (-2\omega_2)^{-r}}\,.
\label{I}
\end{equation}
Let's denote
\begin{equation}
A = 1 - 2 t z + z^2\,,\quad
t = \sin\frac{\vartheta_E}{2}\,,\quad
z = 2 t x\,.
\label{A}
\end{equation}
If $t<1/2$, we can expand
\begin{equation}
A^{-l} = \sum_{k=0}^\infty z^k C^l_k(t)\,,
\label{C}
\end{equation}
where $C^l_k(t)$ are Gegenbauer polynomials;
results for $t>1/2$ can be obtained by analytical continuation.
Expanding also $\Omega/\omega_2$ in $x$,
we obtain from~(\ref{Main})
\begin{eqnarray}
&&I = \sum_{s=0}^\infty \frac{(r)_s}{s!} \xi_-^s J_s\,,
\label{IJ}\\
&&J_s = \sum_{k=0}^\infty \frac{(n_1)_{k+s}}{(n_1+n_2)_{k+s}} (2t)^k C^l_k(t)\,.
\nonumber
\end{eqnarray}

Substituting the explicit form of Gegenbauer polynomials
\[
C^l_k(t) = \sum_{m=0}^{[k/2]} \frac{(-1)^m (l)_{k-m}}{m!\,(k-2m)!} (2t)^{k-2m}
\]
and interchanging the order of summations, we get
\[
J_s = \sum_{m=0}^\infty \sum_{k=2m}^\infty
\frac{(-1)^m (l)_{k-m} (n_1)_{k+s}}{m!\,(k-2m)!\,(n_1+n_2)_{k+s}} (2t)^{2(k-m)}\,.
\]
Now we substitute $k=p+m$ and interchange the order of summations again:
\[
J_s = \sum_{p=0}^\infty (l)_p (2t)^{2p}
\sum_{m=0}^p \frac{(-1)^m (n_1)_{m+p+s}}{m!\,(p-m)!\,(n_1+n_2)_{m+p+s}}\,.
\]
Here the inner sum is a terminating hypergeometric series
\[
\frac{(n_1)_{p+s}}{p!\,(n_1+n_2)_{p+s}}\,
{}_2 F_1\left(\left.
\begin{array}{c}
-p,p+s+n_1\\p+s+n_1+n_2
\end{array}
\right| 1 \right)\,;
\]
expressing it via $\Gamma$ functions, we have
\begin{eqnarray}
&&J_s = \sum_{p=0}^\infty
\frac{(n_1)_{p+s} (n_2)_p (l)_p}{p!\,(n_1+n_2)_{2p+s}} (2t)^{2p}
\nonumber\\
&&{} = \frac{(n_1)_s}{(n_1+n_2)_s}\,
{}_3 F_2 \left( \left.
\begin{array}{c}
s+n_1,n_2,l\\\frac{s+n_1+n_2}{2},\frac{s+n_1+n_2+1}{2}
\end{array}
\right| t^2 \right)\,.
\label{Js}
\end{eqnarray}
Alternatively, interchanging the order of summations in $I$~(\ref{IJ}),
we can write it as
\[
I = \sum_{p=0}^\infty
\frac{(n_1)_p (n_2)_p (l)_p}{p!\,(n_1+n_2)_{2p}} (2t)^{2p}\,
{}_2 F_1 \left( \left.
\begin{array}{c}r,p+n_1\\2p+n_1+n_2\end{array}
\right| \xi_- \right)\,.
\]
Unfortunately, these sums seem not to be expressible
via any well-known special functions.

The result simplifies if $n_1=n_2$.
Using the quadratic transformation
\[
{}_2 F_1 \left( \left.
\begin{array}{c}a,b\\2b\end{array}
\right| z \right) =
\left(1 - \frac{z}{2}\right)^{-a}\,
{}_2 F_1 \left( \left.
\begin{array}{c}\frac{a}{2},\frac{a+1}{2}\\b+\frac{1}{2}\end{array}
\right| \left(\frac{z}{2-z}\right)^2 \right)
\]
and expanding the new hypergeometric function,
we can express $I$ via the Appell function $F_3$~\cite{S:66}:
\begin{eqnarray}
&&\left(\frac{\xi_+}{2}\right)^r I =
\sum_{q=0}^\infty \sum_{p=0}^\infty
\frac{\left(\frac{r}{2}\right)_q \left(\frac{r+1}{2}\right)_q (n_1)_p (l)_p}%
{q!\,p!\,\left(n_1+\frac{1}{2}\right)_{q+p}}
\left(\frac{\xi_-}{\xi_+}\right)^{2q} t^{2p}
\nonumber\\
&&{} = F_3\left(\frac{r+1}{2},l;\frac{r}{2},n_1;n_1+\frac{1}{2};
\frac{\xi_-^2}{\xi_+^2},t^2\right)\,.
\label{F3}
\end{eqnarray}
Here $(r+1)/2+l=n_1+1/2$, and we can reduce this function
to the Appell $F_1$ using
\begin{eqnarray*}
&&F_3(a_1,a_2;b_1,b_2;a_1+a_2;z_1,z_2) ={}\\
&&(1-z_1)^{-b_1}
F_1\left(a_2;b_1,b_2;a_1+a_2;\frac{z_1}{z_1-1},z_2\right)\,.
\end{eqnarray*}
Finally, we arrive at the result
\begin{eqnarray}
&&I(n_1,n_1,n_3;\vartheta;\omega_1,\omega_2) =
I(2n_1,n_3) \left(-2\sqrt{\omega_1\omega_2}\right)^{-r}
\nonumber\\
&&{}\times
F_1\left(l;\frac{r}{2},n_1;n_1+\frac{1}{2};
-\frac{(\omega_1-\omega_2)^2}{4\omega_1\omega_2},\frac{1-\cosh\vartheta}{2}\right)\,.
\label{F1}
\end{eqnarray}
At $\vartheta=0$ it reduces to~(\ref{BBG}), due to the quadratic transformation
\[
{}_2 F_1 \left( \left.
\begin{array}{c}a,b\\2b\end{array}
\right| z \right) =
(1-z)^{-a/2}\,
{}_2 F_1 \left( \left.
\begin{array}{c}\frac{a}{2},b-\frac{a}{2}\\b+\frac{1}{2}\end{array}
\right| - \frac{z^2}{4(1-z)} \right)\,.
\]

\begin{sloppypar}
Another case when the result simplifies is $\omega_1=\omega_2$.
Then the only non-vanishing term in the sum~(\ref{IJ}) is $s=0$,
and from~(\ref{Js}) we obtain the result
\begin{eqnarray}
&&I(n_1,n_2,n_3;\vartheta;\omega,\omega) =
I(n_1+n_2,n_3) (-2\omega)^{-r}
\nonumber\\
&&{}\times {}_3 F_2 \left( \left.
\begin{array}{c}
n_1,n_2,l\\\frac{n_1+n_2}{2},\frac{n_1+n_2+1}{2}
\end{array}
\right| \frac{1-\cosh\vartheta}{2} \right)\,.
\label{F32}
\end{eqnarray}
This form is much simpler than~(\ref{Appell}).
Due to~(\ref{omega0}), we also know $I(n_1,n_2,n_3;\vartheta;\omega,0)$.
If $n_1=n_2$, this hypergeometric function reduces to ${}_2 F_1$,
which agrees with~(\ref{F1}).
\end{sloppypar}

A simple result for $n_1=n_2=1$ can be derived directly from~(\ref{Main}).
Substituting $x=\left[-1+(e^\vartheta+1)z\right]/(e^\vartheta-1)$, we get
\begin{eqnarray}
&&I = (e^{2\vartheta}-1)^{1-2l} e^{\vartheta l} (e^\vartheta-\xi)^{2l-2}
\nonumber\\
&&{}\times\left[F\left(\frac{e^\vartheta}{e^\vartheta+1},y\right)
- F\left(\frac{1}{e^\vartheta+1},y\right)\right]\,,
\nonumber\\
&&F(x,y) = \int_0^x z^{-l} (1-z)^{-l} (1-yz)^{2l-2} dz\,,
\label{F}
\end{eqnarray}
where
\[
y = \frac{(e^\vartheta+1)(1-\xi)}{e^\vartheta-\xi}\,.
\]
Expanding the two brackets in~(\ref{F}) and integrating, we obtain
\begin{eqnarray*}
&&F(x,y) = \frac{x^{1-l}}{1-l}
\sum_{n,m=0}^\infty
\frac{(l)_n (2-2l)_m (1-l)_{n+m}}{n!\,m!\,(2-l)_{n+m}}
x^n (xy)^m\\
&&{} = \frac{x^{1-l}}{1-l}
F_1(1-l;l,2-2l;2-l;x,xy)\,.
\end{eqnarray*}
This Appell function can be reduced using
\begin{eqnarray*}
&&F_1(a;b_1,b_2;b_1+b_2;z_1,z_2) = {}\\
&&(1-z_2)^{-a}\,{}_2 F_1 \left( \left.
\begin{array}{c}a,b_1\\b_1+b_2\end{array}
\right| \frac{z_1-z_2}{1-z_2} \right)\,,
\end{eqnarray*}
and so
\[
F(x,y) = \frac{(x^{-1}-y)^{l-1}}{1-l}\,
{}_2 F_1 \left( \left.
\begin{array}{c}1-l,l\\2-l\end{array}
\right| \frac{1-y}{x^{-1}-y} \right)\,.
\]
At last, we arrive at
\begin{eqnarray}
&&I(1,1,n;\vartheta;\omega_1,\omega_2) = I(2,n)
\left(-2\sqrt{\omega_1\omega_2}\right)^{-r}
\nonumber\\
&&{}\times\frac{1}{l-1} (e^\vartheta-e^{-\vartheta})^{-l}
(e^{\vartheta/2}\xi^{-1/2}-e^{-\vartheta/2}\xi^{1/2})^{l-1}
\nonumber\\
&&{}\times\Biggl[ (e^{-\vartheta}\xi)^{(1-l)/2}\,
{}_2 F_1 \left( \left.
\begin{array}{c}1-l,l\\2-l\end{array}
\right| \frac{\xi-e^{-\vartheta}}{e^\vartheta-e^{-\vartheta}} \right)
\nonumber\\
&&\qquad{} - (e^{-\vartheta}\xi)^{(l-1)/2}\,
{}_2 F_1 \left( \left.
\begin{array}{c}1-l,l\\2-l\end{array}
\right| \frac{e^\vartheta-\xi^{-1}}{e^\vartheta-e^{-\vartheta}} \right)
\Biggr]\,.
\label{F21}
\end{eqnarray}
At $\omega_1=\omega_2$ this should be equivalent to~(\ref{F32}).

\section{Expansions in $\varepsilon$}
\label{Exp}

The hypergeometric $\omega_1=\omega_2$ result~(\ref{F32})
can be expanded in $\varepsilon$.
Expansion of such hypergeometric functions was considered
in a number of papers~\cite{K}.
It can be performed automatically
using the \texttt{Mathematica} package \texttt{HypExp}~\cite{HM}.
Coefficients are expressed via harmonic polylogarithms~\cite{RV},
and can be simplified by the package \texttt{HPL}~\cite{M}.
It is better to work in Euclidean space,
where arguments of logarithms and polylogarithms are away from branch cuts,
and to do analytical continuation to Minkowski space at the end.

All integrals reduce to 3 master ones.
Their expansions have the form
\begin{eqnarray}
&&\frac{I(1+2l_1\varepsilon,1+2l_2\varepsilon,1+l_3\varepsilon;\vartheta;\omega,\omega)}%
{I(2+2(l_1+l_2)\varepsilon,1+l_3\varepsilon)
(-2\omega)^{-2(1+l_1+l_2+l_3)\varepsilon}}
\nonumber\\
&&{}=\frac{1}{\sinh\vartheta}
\left[ \vartheta + a_1(\vartheta) \varepsilon + a_2(\vartheta) \varepsilon^2 + \cdots \right]\,,
\label{expa}\\
&&\frac{I(1+2l_1\varepsilon,2l_2\varepsilon,1+l_3\varepsilon;\vartheta;\omega,\omega)}%
{I(2+2(l_1+l_2)\varepsilon,1+l_3\varepsilon)
(-2\omega)^{1-2(1+l_1+l_2+l_3)\varepsilon}}
\nonumber\\
&&{}=1 - 2 l_2 \varepsilon
\left[\tau\vartheta + b_1(\vartheta) \varepsilon + b_2(\vartheta) \varepsilon^2 + \cdots \right]\,,
\label{expb}
\end{eqnarray}
where $\tau=\tanh(\vartheta/2)$ (expansion of
$I(2l_1\varepsilon,1+2l_2\varepsilon,1+l_3\varepsilon;\vartheta;\omega,\omega)$
is given by the formula symmetric to~(\ref{expb})).
At $l_2=0$, the right-hand side of~(\ref{expb}) is exactly 1;
$a_n(\vartheta)$ are odd functions ($\sim\vartheta^3$ at $\vartheta\to0$);
$b_n(\vartheta)$ are even functions ($\sim\vartheta^2$ at $\vartheta\to0$).

The first corrections are
\begin{eqnarray}
&&a_1(\vartheta) = c_1(\vartheta) + 2 l_{12} \vartheta\,,
\label{a1}\\
&&b_1(\vartheta) = \tau c_1(\vartheta) - \frac{1}{2} j' \vartheta^2\,,
\label{b1}\\
&&c_1 = - 2 l_{12} L_{2a} - j L_{2b}\,,
\nonumber
\end{eqnarray}
where $l_{12}=l_1+l_2$, $j_3=1+l_3$, $j=j_3-l_{12}$, $j'=j_3-l_1+l_2$,
\begin{eqnarray*}
&&L_{2a} = \Li2(\tau) - \Li2(-\tau)\,,\\
&&L_{2b} = \Li2\left(\frac{1+\tau}{2}\right) - \Li2\left(\frac{1-\tau}{2}\right)
- L \vartheta\,,\\
&&L = - \frac{1}{2} \log \frac{1-\tau^2}{4}
= \log \left(2 \cosh\frac{\vartheta}{2} \right)\,.
\end{eqnarray*}
Two dilogarithms are not independent:
\[
\Li2\left(\frac{1+\tau}{2}\right) + \Li2\left(\frac{1-\tau}{2}\right)
= - L^2 + \frac{\vartheta^2}{4} + \frac{\pi^2}{6}\,;
\]
$L_{2b}$ is written here in the manifestly odd form.

The second corrections are
\begin{eqnarray}
&&a_2(\vartheta) = c_2(\vartheta) - 4 l_{12}^2 L_{2a} - 2 l_{12} j L_{2b}\,,
\label{a2}\\
&&b_2(\vartheta) = \tau c_2(\vartheta)
+ 4 j' j_3 L_{3d}
\nonumber\\
&&\quad{} - 2 \left[(3l_1+l_2) j_3 - l_1^2 + l_2^2\right] L_{3e}
\nonumber\\
&&\quad{} - j_3 \vartheta \left[ 4 l_1 L_{2a}
+ (j_3-3l_1+l_2) L_{2b} + j' L \vartheta \right]\,,
\label{b2}\\
&&c_2(\vartheta) = 4 l_{12} j_3 L_{3a} + 2 j j_3 L_{3b} - 2 l_{12} j L_{3c}
\nonumber\\
&&\quad{} - \frac{1}{12} (j_3^2 - 5 l_{12} j_3 + 8 l_1 l_2) \vartheta^3\,,
\nonumber
\end{eqnarray}
where
\begin{eqnarray*}
&&L_{3a} = \Li3(\tau) - \Li3(-\tau)\,,\\
&&L_{3b} = \Li3\left(\frac{1+\tau}{2}\right) - \Li3\left(\frac{1-\tau}{2}\right)
+ \frac{\vartheta}{2} \left( L^2 - \frac{\pi^2}{6} \right)\,,\\
&&L_{3c} = \Li3\left(\frac{2\tau}{1+\tau}\right) - \Li3\left(\frac{-2\tau}{1-\tau}\right)\,,\\
&&L_{3d} = \Li3\left(\frac{1+\tau}{2}\right) + \Li3\left(\frac{1-\tau}{2}\right)\\
&&\quad{} - \frac{1}{3} L^3 + \frac{\pi^2}{6} L - \frac{7}{4} \zeta(3)\,,\\
&&L_{3e} = \Li3\left(\frac{2\tau}{1+\tau}\right) + \Li3\left(\frac{-2\tau}{1-\tau}\right)
\end{eqnarray*}
($L_{3d}$ can be written via a single trilogarithm $\Li3(-(1\pm\tau)/(1\mp\tau))$.

The next correction can be derived by \texttt{HypExp};
it contains harmonic polylogarithms which cannot be reduced
to ordinary polylogarithms.

If $n_1=n_2=1$, $I$~(\ref{I}) depends on $n_3$ and $\varepsilon$
only via the product $j_3\varepsilon$, see~(\ref{Main}).
The coefficients~(\ref{a1}), (\ref{a2}) satisfy this requirement,
and coincide with the expansion of~(\ref{F21}) at $\omega_1=\omega_2$.

At small $\vartheta$,
keeping only the leading $\vartheta^3$ terms in $a_{1,2}(\vartheta)$
and the leading $\vartheta^2$ terms in $b_{1,2}(\vartheta)$,
we reproduce the $\vartheta^2$ term in~(\ref{F32})
expanded in $\varepsilon$.

\section{Conclusion}
\label{Conc}

In the single-scale case $\omega_1=\omega_2$,
the vertex diagram with arbitrary powers of all denominators
is given by the surprisingly simple formula~(\ref{F32}).
The cases $\omega_{1,2}=0$ reduce to this one~(\ref{omega0}).
Expansions in $\varepsilon$ has been derived in Sect.~\ref{Exp}.
If $\omega_1\ne\omega_2$, we were able to obtain
relatively simple formulas for $n_1=n_2$~(\ref{F1})
and especially $n_1=n_2=1$~(\ref{F21}).

The integral~(\ref{Definition}) with $n_1=n_2=n_3=1$ was considered
in~\cite{B:00} for general $\omega_{1,2}$
(in fact, even a more general integral with a nonzero mass $m$
of the light line was calculated up to $\mathcal{O}(\varepsilon^0)$).
Our $a_1$~(\ref{a1}) with $l_1=l_2=l_3=0$ reproduces
the formula~(37) from this paper.

Integrals with $n_1=n_2=n_3=1$; $n_1=n_2=1$, $n_3=1+\varepsilon$;
$n_1=1+2\varepsilon$, $n_2=n_3=1$ were considered in the Appendix of~\cite{K:09}
(the formulas~(69), (71), (72)).
It is difficult to make sense of these formulas,
because all integrals in the left-hand sides of equations in this Appendix
are equal to 0, and the right-hand sides have dimensionality
different from the left-hand sides.
Evidently, the author assumed some different integrals in the left-hand sides,
and some dimensional factors in the right-hand sides,
but it seems impossible to guess their real meaning.
In particular, these formulas cannot be interpreted as our integrals
with $\omega_1=\omega_2$ with appropriate powers of $-2\omega$ inserted
into the right-hand sides,
because already their values at $\vartheta=0$ differ.
It would be interesting to compare the formulas~(69), (71)
(containing ${}_2F_1$ hypergeometric functions) and~(72)
(containing the Appell $F_1$ function) to our results.
The $\mathcal{O}(\varepsilon^0)$ term in~(69) contains dilogarithms
similar to those in $a_1$~(\ref{a1}) (at $l_1=l_2=l_3=0$),
but the leading $\mathcal{O}(\varepsilon^{-1})$ terms differ.

\begin{acknowledgement}
We thank T.~Huber and D.~Ma\^{\i}tre for providing updated versions
of \texttt{HypExp} and \texttt{HPL} with bug fixes,
and to M.Yu.~Kalmykov for useful discussions.
The work of AGG was supported by the BMBF through grant No.\ 05H09VKE.
\end{acknowledgement}

\end{document}